# Exploring the Impact of Configurational Entropy on the Design and Development of CoNi-Based Superalloys for Sustainable Applications

Ahad Mohammadzadeh[a, b*], Akbar Heidarzadeh[c], Hailey Becker[d], Jorge Valilla Robles[a,e], Alberto Meza[a], Manuel Avella[a], Miguel A. Monclús[a], Damien Tourret[a], José Manuel Torralba[a, e]

[a] Imdea Materials Institute, Calle Eric Kandel, 2, 28906, Getafe, Madrid, Spain

[b] Department of Materials Engineering, Faculty of Engineering, University of Maragheh, Maragheh, P.O. Box 83111-55181, Iran

[c] Department of Materials Engineering, Azarbaijan Shahid Madani University, Tabriz 53714-161, Iran

[d] Department of Chemical Engineering and Materials Science, Michigan State University, USA

[e] Universidad Carlos III de Madrid, Av. Universidad 30, 28911 Leganés

*Corresponding Author. E-mail address: ahad.mohammadzadeh@imdea.org

**Abstract**

A comprehensive literature review on recently rediscovered Co- and/or CoNi-based superalloys, strengthened by the γ' phase, revealed a relationship between the configurational entropy of the system and the γ' solvus temperature. This study was conducted on a high Cr CoNi-based superalloy system with high configurational entropy to test our hypothesis based on the sustainable metallurgy framework. Thermodynamic calculations were performed to design the chemical compositions, followed by vacuum casting and heat treatments to produce the desired alloys. The microstructures were characterized using a scanning electron microscope, electron backscattered diffraction, transmission electron microscope, and differential thermal analysis. Microhardness and nanoindentation tests were employed to measure the mechanical properties. The results showed that both the configurational entropy and the type of alloying elements determine the final high-temperature performance of the alloys. We found that to enhance the higher γ' solvus temperature, the configurational entropy should be increased by adding γ' stabilizing elements. The microstructural and mechanical characteristics of the designed alloys before and after heat treatments are discussed in detail. The outcome of this study is beneficial for developing cobalt-based high-entropy superalloys with appropriate processing windows and freezing ranges for *advanced sustainable manufacturing* purposes, such as using powder bed fusion technologies.

**1 Introduction**

Improving energy efficiency and effectively capturing and storing greenhouse gas emissions, such as $CO_2$, are crucial responsibilities for all industrial sectors. These objectives necessitate the development of new technologies, with the ultimate goal of achieving zero-emission power plants. Achieving this goal relies heavily on the availability of suitable materials, such as superalloys, and the exploration of new processing methods [1,2]. Superalloys have garnered significant attention in the aerospace and power generation industries due to their remarkable combination of mechanical and chemical properties [2–5]. Today, they are successful in addressing the pressing demands for durability and strength in machinery and systems that were unimaginable a hundred years ago. Among the various types of superalloys, Ni-based superalloys stand out as the earliest and most extensively developed family. These alloys possess a two-phase microstructure comprising a gamma (γ) matrix and a strengthening phase known as gamma prime (γ′), characterized by $L1_2$ crystallography. Ni-based superalloys have found extensive use in modern aircraft engines, constituting approximately 40-50 wt. % of their composition [6–8]. However, the progress of mature Ni superalloys is limited by the γ′ solvus temperature, which approaches the Ni melting point. This limitation hampers efforts to enhance their sustainability by improving energy efficiency and reducing $CO_2$ emissions. Fortunately, the introduction of novel $L1_2$-strengthened Co-based superalloys, currently in the early stages of development, offers a viable alternative to some Ni-based superalloys. These Co-based superalloys exhibit higher melting points, ranging from 50-150 °C higher than Ni superalloys [9–12].

Since the rediscovery of the γ'-$Co_3$(Al, W) phase in 2006 [13], significant efforts have been dedicated to comprehending the thermodynamic, physical, mechanical, and environmental properties of γ'-strengthened Co- or CoNi-based superalloys [14–20]. Furthermore, researchers have successfully developed novel Co- or CoNi-based superalloys with an increased γ' solvus temperature and higher γ' volume fraction [21–24]. Based on the historical development of superalloys [25] and a comprehensive analysis of data published in the literature [26–37], a relationship has been identified between the entropy of mixing ($\Delta S_{mix}$) and the γ' solvus temperature in Co- or CoNi-based superalloys, as illustrated in Fig. 1. For instance, enhancing $\Delta S_{mix}$ from 0.613R (four elements) to 1.298R (six elements) raises the γ' solvus temperature from 870 °C to 1128 °C [31]. The figure suggests that designing novel alloys based on the high entropy concept [38], i.e., producing high entropy alloys (HEAs), holds promise in overcoming the limitations of the γ' solvus temperature in Co-based superalloy applications. Building upon the successful platform established for Ni-based superalloys, the development of Co-based superalloys/HEAs can significantly accelerate the advancement of high-temperature materials. Fig. 1 also indicates that, even at the same level of entropy, some alloys within each category exhibit higher temperature resistance than others, underscoring the importance of selecting appropriate elements and their concentrations. Therefore, gaining a fundamental understanding of the effects of different elements is crucial to the development of suitable Co- or CoNi-based high entropy superalloys (HESAs) with exceptional properties, necessitating further research.

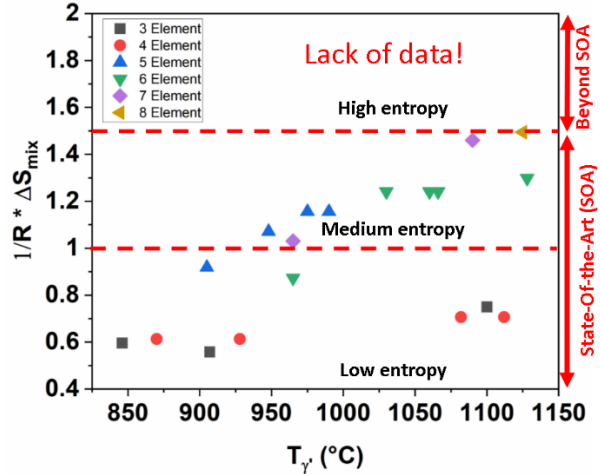

Fig. 1. Relationship between ΔS$_{mix}$ and γ′ solvus temperature in Co- or CoNi-based superalloys, data from literature [26–37].

A promising CoNi-based superalloy system, composed of six components (Co-Ni-Al-Ti-Ta-W), has been recognized for its higher γ' solvus temperature without the formation of topologically close-packed (TCP) phases at high temperatures [29,39]. In addition to alloy design considerations, such as increasing oxidation and corrosion resistance, reducing density (critical for aerospace applications) is also important. Density reduction can be achieved by replacing W with alternative alloying elements [40–44]. Li et al. [41] investigated the impact of substituting W with γ' former elements in a Co-20Ni-7Al-8W-1Ta-4Ti (all in at. %, throughout the text, the chemical composition is reported in at %.) superalloy. The alloy underwent an aging process at 1000°C for 1000 hours. By employing the diffusion-multiple approach and conducting thermodynamic calculations, they found that excessive additions of Al, Ta, Mo, and Nb led to the formation of different phases, whereas the addition of Ti did not induce detrimental phase formation. W played a crucial role in preserving the thermal stability of the γ' phase, while Ti and Ta effectively enhanced this stability. In comparison, Mo exhibited lower efficiency compared to W. The researchers suggest that Ti shows promise as a lightweight alternative to W, and the adverse effect on γ' strength can be mitigated by moderate additions of Ta and Nb.

For improved oxidation and corrosion resistance, the addition of Cr alongside Al becomes necessary in multicomponent Co- or CoNi-based superalloys, leading to the development of high-Cr CoNi-based superalloys [27,45–47]. For instance, Zhuang et al. [27] employed the calculation of phase diagrams (CALPHAD) method to design an alloy with high-temperature oxidation and corrosion resistance. Their composition, Co-30Ni-8Al-4Ti-2W-1Ta-14Cr, followed a strategy similar to commercially available wrought Ni-based superalloys, where Al and Al+Cr content was set equal to or higher than 4.5 at. % and 21 at. %, respectively. Liu et al. [28] utilized a machine learning method to optimize an alloy, resulting in an ideal composition of 2 at. % Cr and 12 at. % Al, with a γ' solvus temperature of 1266.5 °C and a γ' volume fraction of 74.5% after aging at 1000 °C for 1000 h. They emphasized that the desired oxidation resistance in the developed CoNi-based superalloy is linked to the formation of a protective alumina layer at 1000 °C.

Additionally, Zhuang et al. [23] investigated the effects of alloying elements in a high Cr CoNi-based superalloy. They reported that Cr reduces the volume fraction of γ' phase, while Al, Ti, Nb, and W additions increase it. The introduction of Cr and Nb leads to smaller γ' sizes and slower coarsening, while elements like Mo and Ti result in larger γ' sizes and faster coarsening. Furthermore, Cr, Mo, and Ni enhance the spherical morphology of γ' precipitates, while Co addition or the presence of Nb, Ta, Al, and Ti promotes a more cuboidal or elongated shape. Li et al. [42] reported that Cr can mitigate the adverse impact of W on the γ' volume fraction in CoNi superalloys by modifying the partitioning characteristics of Al and Ti. Recently, Cao et al. [48] introduced a novel multicomponent W-free Co-rich HEA containing 10 at. % Cr, reporting an increased γ' solvus temperature up to 1125 °C, surpassing the values reported in the literature for W-free Co-based superalloys.

In this study, we investigate the impact of $\Delta S_{mix}$ on the γ' solvus temperature in high Cr CoNi-based superalloys, taking into account existing findings from the literature [21,27–29]. Using thermodynamic calculations based on the CALPHAD method, we designed new CoNi-based HESAs by considering alloying elements with high-temperature capability and low density, while targeting $\Delta S_{mix}$ values higher than 1.5R. The results of this research introduce a novel design strategy that combines high $\Delta S_{mix}$ and the CALPHAD method, offering potential advancements in the field of high Cr CoNi-based superalloys.

## 2 Materials and methods

*2.1 CALPHAD calculations and alloy design*

We conducted calculations for multicomponent phase diagrams using the software ThermoCalc and the Ni-based superalloy database TCNI8. In the initial round of calculations, which involved the application of the lever rule, Gulliver-Scheil method, and multicomponent phase diagram analysis, we encountered inaccuracies in the results. These inaccuracies can be attributed to the limited information in the database and the novelty of the alloys being investigated. After conducting a thorough literature review of microstructures in similar or closely related alloys, we narrowed down the number of considered phases to those deemed most relevant (see the results section). The design criteria taken into consideration were as follows: a $\Delta S_{mix}$ value higher than 1.5R, a higher gamma prime solvus temperature, a lower density, and the prevention of TCP phase formation. The ideal $\Delta S_{mix}$ value was calculated as follows:

$$\Delta S_{mix} = -R \sum x_i \ln x_i \quad (1)$$

where R is the gas constant and $x_i$ represents the mole fraction of the i[th] element, we performed calculations using a septenary alloy composition of 41Co-30Ni-8Al-4Ti-2W-1Ta-14Cr [20]. This particular alloy demonstrated a $\Delta S_{mix}$ value of 1.457R. To increase $\Delta S_{mix}$, we made modifications to the alloy composition. Ni content was raised to 35 at. % at the expense of Co and Ta concentration was increased to 2 at. % at the expense of Cr. Additionally, we introduced V as a replacement for a portion of the chromium. During the design phase, our objective was to optimize the quantities of all elements based on the available literature to attain their respective optimal values. Once we optimized the element ratios, we explored the possibility of further enhancing $\Delta S_{mix}$ by incorporating additional elements, specifically through the inclusion of V. As a result of our calculations, we narrowed down our selection to three CoNi-based HESAs named CNS1, CNS2, and CNS3, each exhibiting septenary and octonary compositions (see Table 1).

Table 1. Nominal (labelled No.) and experimental (labelled Ex., measured by EDS point analysis) chemical composition of the homogenized/air-cooled samples in conjunction with their $\Delta S_{mix}$.

| Sample codes | | Chemical composition (at. %) | | | | | | | |
|---|---|---|---|---|---|---|---|---|---|
| | | Co | Ni | Al | Ti | V | W | Ta | Cr |
| CNS1 | No. | 36 | 35 | 8 | 4 | 0 | 2 | 2 | 13 |
| | Ex. | 36.10±0.15 | 34.45±0.32 | 7.92±0.09 | 4.35±0.08 | 0 | 1.71±0.1 | 2.21±0.1 | 13.26±0.21 |
| CNS2 | No. | 36 | 35 | 8 | 4 | 4 | 2 | 2 | 9 |
| | Ex. | 35.97±0.4 | 34.49±0.42 | 7.68±0.3 | 4.36±0.23 | 4.27±0.06 | 1.87±0.12 | 2.06±0.08 | 9.31±0.05 |
| CNS3 | No. | 40 | 30 | 8 | 4 | 2 | 2 | 2 | 12 |
| | Ex. | 40.15±0.27 | 29.33±0.26 | 7.81±0.26 | 4.1±0.01 | 2.12±0.02 | 2.06±0.11 | 2.05±0.10 | 12.4±0.04 |

*2.2 Experimental procedure*

To evaluate the accuracy of the thermodynamic calculations, the alloys CNS1, CNS2, and CNS3 (refer to Table 1) were fabricated through vacuum arc remelting (VAR) under a high-purity Ar atmosphere. To compensate for Al evaporation during melting, the mass of Al was adjusted to be 5% higher than the target values. The alloys underwent multiple cycles of inversion, melting, and remelting to ensure homogeneous chemical compositions. Following casting, the alloys were homogenized at 1250 °C for 24 hours, followed by cooling in air and in a furnace. Aging was carried out on the air-cooled samples at 900 °C for 24 hours and subsequently quenched in water.

Characterization of the microstructures in the as-cast, homogenized, and aged samples was conducted using an FEI Helios Nanolab 600i scanning electron microscope (SEM) equipped with energy-dispersive X-ray spectroscopy (EDS) and Nordlys electron backscatter diffraction (EBSD) detectors. Prior to EBSD analysis, the samples were mechanically ground and polished using a series of diamond pastes, gradually reducing the grit size to 1 μm, followed by a final polishing step using an oxide particle suspension (OPS) with a particle size of 0.04 μm to improve the surface quality. Kalling's number 2 solution was used for etching the samples to reveal the γ and γ' phases in the homogenized and aged conditions for 20 seconds. The area fraction of the γ' phase in SEM images was determined using ImageJ software [49]. EBSD data analysis was performed using AZtecHKL acquisition software. For further microstructural details, such as morphology and elemental partitioning in the γ and γ' phases, an FEI Talos F200X 200keV field emission scanning/transmission electron microscope (STEM) was employed. The STEM samples were thinned using grinding paper to approximately 50 μm and then punched into 3 mm disks. The final thinning was accomplished using a Struers TenuPol-5 electropolishing unit with a solution composed of 40 mL acetic acid, 80 mL perchloric acid, and 880 mL ethanol at -25 °C and 22.5 V.

Differential thermal analysis (DTA) was performed on the aged samples using a Setsys Evolution TGA & DTA/DSC Setaram instrument to determine the solidus, liquidus, and γ' solvus temperatures. The DTA experiments were conducted in an alumina crucible under an Ar atmosphere with a heating rate of 10 °C/min.

The mechanical properties of the as-cast, homogenized, and aged samples were assessed using Vickers hardness and nanoindentation tests. Vickers hardness was measured using an INNOVATEST instrument with a 1 kg load and a dwell time of 15 s at room temperature. Five hardness measurements were taken per sample, and the average values were recorded. For the as-cast and homogenized samples, nanoindentation arrays were created using a Hysitron TI 950 Triboindenter equipped with a diamond Berkovich indenter. Arrays of 22 × 22 indents were performed with a spacing of 3.5 μm, employing a 0.1 s linear ramp to a peak force of 2 mN, a 0.1 s hold time, and a 0.1 s linear unloading ramp, resulting in maximum depths of 100 nm. For the aged samples, nanoindentation tests were conducted in load-controlled mode using a load function comprising a 10 s load, 5 s hold, and 2 s unload, with a maximum load of 500 mN. The reported values for nanoindentation hardness ($H_{ind}$) and reduced modulus ($E_r$) represented the average of at least 9 indents and were obtained using the Oliver and Pharr method [50] by analyzing the force-displacement curves generated during the tests.

## 3 Results

### 3.1 CALPHAD calculations

As stated in the method section, the final CALPHAD calculations involved a reduced set of phases that were considered the most relevant. Consequently, secondary phases such as $D0_{19}$, $D0_{24}$, and Laves phases were excluded since they have not been reported in this series of CoNi-based superalloys [23,27,45]. Fig. 2 illustrates the resulting isopleth sections (lever rule) for the designed CoNi-based HESAs. The isopleth sections were calculated for compositions XCo-YNi-8Al-4Ti-2Ta-2W-13Cr (CNS1), XCo-YNi-8Al-4Ti-4V-2Ta-2W-9Cr (CNS2), and XCo-YNi-8Al-4Ti-2V-2Ta-2W-12Cr (CNS3), where the Ni content was varied from 30 at. % to 40 at. %, with the expense of Co, and plotted against a temperature range of 750 °C to 1500 °C.

The isopleth section of phase diagrams reveals extensive regions of γ and γ+γ′, suggesting suitable areas for solutionizing and aging heat treatments. To increase $\Delta S_{mix}$, a balance between Co and Ni is necessary, especially in a septenary system. Therefore, the Ni content for the CNS1 alloy was adjusted to 35 at. % (Fig. 2a). Notably, at a Ni content of 35 at. %, the formation of β and σ phases at temperatures higher than 900 °C can be avoided. The CNS2 alloy was developed to incorporate eight elements (in order to increase $\Delta S_{mix}$), with the addition of V at the expense of Cr.

Consequently, the addition of V raised the ΔS$_{mix}$ level to 1.568R (Table 2). It is important to mention that the influence of V as a lightweight γ′ stabilizer [51] on the microstructural characteristics of this series of CoNi-based superalloys has not yet been investigated.

Table 2. Nominal and experimental values of ΔS$_{mix}$ for the developed CoNi-based HESAs.

| Sample codes | | $\frac{1}{R}\Delta S_{mix}^{Co}$ | $\frac{1}{R}\Delta S_{mix}^{Ni}$ | $\frac{1}{R}\Delta S_{mix}^{Al}$ | $\frac{1}{R}\Delta S_{mix}^{Ti}$ | $\frac{1}{R}\Delta S_{mix}^{V}$ | $\frac{1}{R}\Delta S_{mix}^{W}$ | $\frac{1}{R}\Delta S_{mix}^{Ta}$ | $\frac{1}{R}\Delta S_{mix}^{Cr}$ | $\Delta S_{mix}$ |
|---|---|---|---|---|---|---|---|---|---|---|
| CNS1 | No. | 0.368 | 0.367 | 0.202 | 0.129 | 0 | 0.078 | 0.078 | 0.265 | 1.488R |
|  | Ex. | 0.368 | 0.367 | 0.201 | 0.136 | 0 | 0.070 | 0.084 | 0.068 | 1.494R |
| CNS2 | No. | 0.368 | 0.367 | 0.202 | 0.129 | 0.129 | 0.078 | 0.078 | 0.217 | 1.568R |
|  | Ex. | 0.368 | 0.367 | 0.197 | 0.137 | 0.135 | 0.074 | 0.080 | 0.221 | 1.579R |
| CNS3 | No. | 0.367 | 0.361 | 0.202 | 0.129 | 0.078 | 0.078 | 0.078 | 0.254 | 1.548R |
|  | Ex. | 0.366 | 0.360 | 0.199 | 0.131 | 0.082 | 0.080 | 0.080 | 0.259 | 1.556R |

In the case of octonary systems where the formation of β and µ phases is possible, the Ni content for the CNS2 alloy was set at 35 at. % (Fig. 2b). From Fig. 2b, it can be observed that the β phase region shifted to lower Ni content due to the decrease in Cr and the addition of V. Furthermore, the isopleth section of CNS2 alloy indicates that it may be possible to avoid µ phase formation at temperatures higher than 850 °C. The CNS3 alloy, which contains a higher amount of γ stabilizer, was modified to have a ΔS$_{mix}$ value of 1.548R. Additionally, the alloy has an increased Co content of 40 at. % and a Ni of 30 at. %. This modified alloy was specifically designed to investigate the impact of nickel (decreasing γ′ stabilizer) on microstructural characteristics and phase transformations. From Fig. 2c, the possibility of β and µ phases formation (as well as σ phase at temperatures lower than 750 °C, not represented in Fig. 2) appears to be higher. Referring to the isopleth section for the CNS3 alloy (Fig. 2c), there is a narrow temperature range for the possible formation of the β phase (~940 °C to ~1130 °C), which could potentially be avoided through homogenization treatment. Both σ and µ phases could be avoided at temperatures higher than 900 °C.

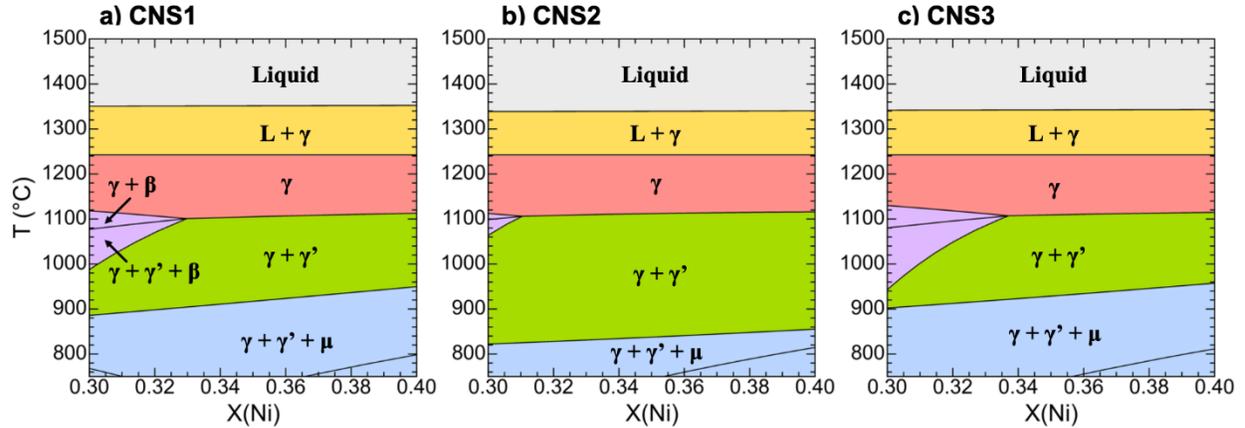

Fig. 2. (a-c) Phase diagram isopleth sections showing the effect of Ni alloy content at the expense of Co for the developed CoNi-based HESAs.

*3.1 Validation of the CALPHAD calculations*

A typical SEM backscattered image of the as-cast CNS1 alloy is depicted in Fig. 3a, revealing a dendritic structure with evidence of interdendritic microsegregation (indicated by white arrows). EBSD and EDS analyses were performed on the as-cast alloys to identify the phases present in the microstructure. The EBSD phase map (Fig. 3b) confirmed the existence of a second phase with a body-centered cubic (BCC) crystallographic structure. Considering the EDS mapping analysis (Fig. 3c), the BCC phase was identified as an Al- and Ti-rich intermetallic within the segregation regions. Referring to the isopleth section of CNS1 (Fig. 2a), the possible phase that could form during solidification is the β phase, which is consistent with the EBSD and EDS results.

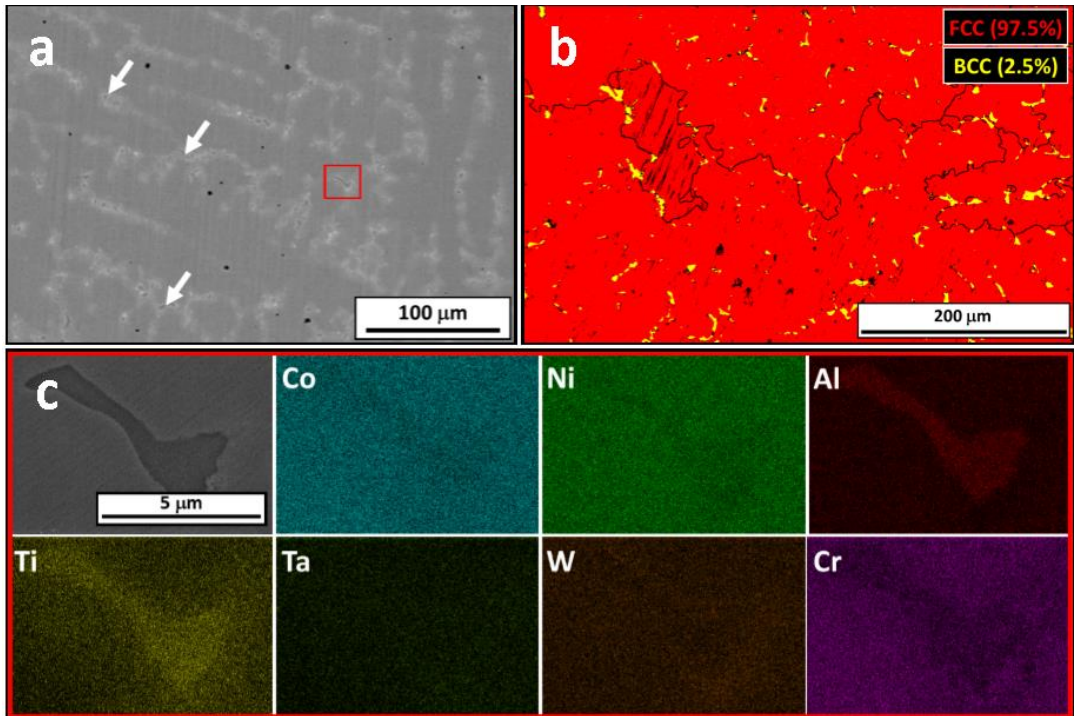

**Fig. 3.** Microstructural characterization of the as-cast CNS1 alloy: (a) SEM backscattered image, (b) EBSD phase map, and (c) EDS mapping analysis of the area indicated by the red rectangle in (a). In (b), the high-angle grain boundaries (θ≥15°) are depicted by black lines. The EBSD scan had a step size of 0.4 μm.

In order to homogenize the chemical composition, dissolution of the β phase, and obtain a uniform microstructure, the as-cast alloys underwent solutionizing at 1250 °C for 24 hours, followed by air and furnace cooling. Fig. 4 presents the SEM micrographs of the solutionized samples, demonstrating a more refined distribution of γ/γ′ in the air-cooled condition (Figs. 4a-c) compared to the furnace-cooled condition (Figs. 4d-f). The presence of coarser γ′ after the solutionizing heat treatment makes it unsuitable for subsequent aging heat treatment, thus the air-cooled samples were selected for further analysis. Supplementary material **S1**, comprising EBSD phase maps of the air-cooled samples, confirms the successful solutionizing process as the main matrix phase exhibited an FCC structure, with only a minor amount of β phase (≤ 0.2%) showing a BCC structure. The chemical compositions of the solutionized samples (Table 1) closely matched the nominal values based on experimental measurements. Additionally, the calculated ΔS$_{mix}$ values, based on both the nominal and experimental values, are summarized in Table 2, highlighting higher ΔS$_{mix}$ values compared to nominal values.

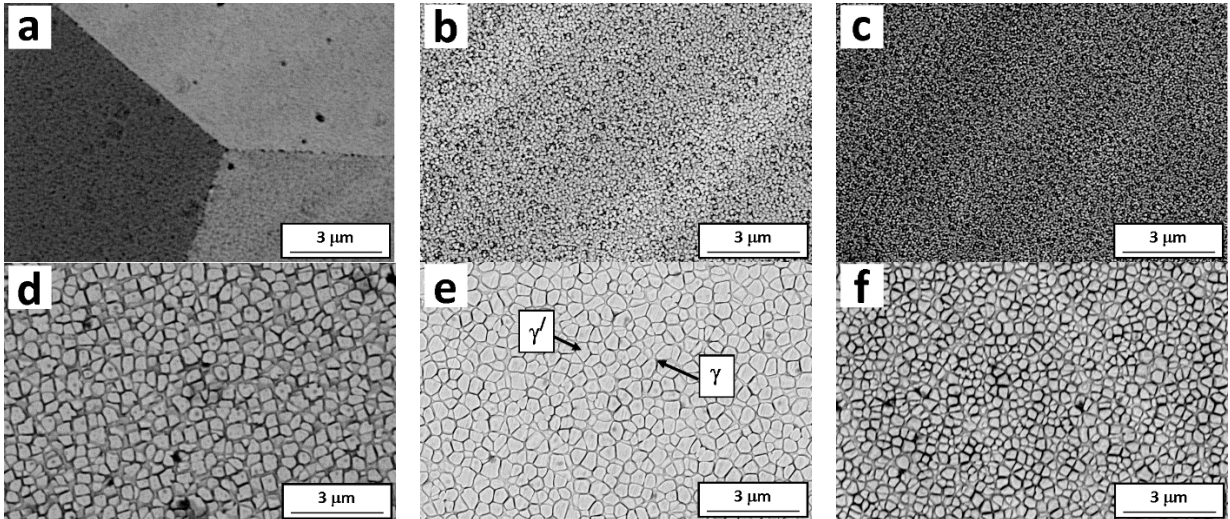

Fig. 4. SEM micrographs of the solutionized samples at 1250 °C for 24 hours, followed by air-cooling (a-c) and furnace-cooling (d-f) for CNS1 (a and d), CNS2 (b and e), and CNS3 (c abd f) alloys, respectively.

*3.2 Aging*

The DTA curves of the aged samples (Fig. 5) were utilized to determine the actual values of the solidus (T$_S$), liquidus (T$_L$), and γ' solvus (T$_{γ'}$) temperatures. The actual and predicted values of T$_S$, T$_L$, and T$_{γ'}$, along with the corresponding freezing range and hot processing window for the aged samples, are summarized in Table 3. Based on Fig. 5 and Table 3, the CNS2 sample exhibited a higher T$_{γ'}$ temperature (1165 °C) compared to the CNS1 (1150 °C) and CNS3 (1134 °C) samples. Furthermore, a hot processing window between 120 °C and 141 °C and the freezing range between 61 °C and 71 °C (Table 3) indicate the significant potential of the developed alloys for *sustainable manufacturing purposes*, such as powder bed fusion techniques.

Table 3 shows a comparison between the predicted values for T$_S$, T$_L$, and T$_{γ'}$ and the measured values for aged samples. The predictions for T$_L$ demonstrate a reasonable agreement with the DTA results. However, it is evident that the predictions for $T_S$ and $T_{\gamma'}$ do not show a consistent alignment with the experimental findings.

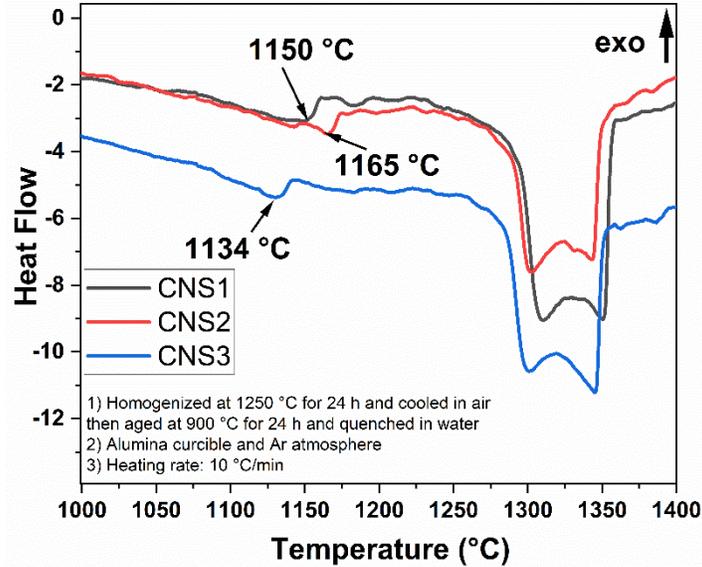

Fig. 5. DTA curves of aged samples.

Table 3. Predicted and measured (actual) values of the $T_S$, $T_L$, and $T_{\gamma'}$ temperatures of the aged samples, along with the corresponding freezing range and hot processing window.

| Sample code | Calculation method | $T_S$ (°C) | $T_L$ (°C) | $T_{\gamma'}$ (°C) | Freezing range (°C) | Hot processing window (°C) |
|---|---|---|---|---|---|---|
| CNS1 | Actual | 1285 | 1351 | 1150 | 66 | 135 |
|  | Predicted | 1229 | 1351 | 1116 | 122 | 113 |
| CNS2 | Actual | 1285 | 1346 | 1165 | 61 | 120 |
|  | Predicted | 1230 | 1339 | 1123 | 109 | 107 |
| CNS3 | Actual | 1275 | 1334 | 1134 | 71 | 141 |
|  | Predicted | 1241 | 1342 | 1080 | 101 | 161 |

Fig. 6 depicts the SEM micrographs of the aged samples, revealing the presence of two phases: γ (black) and γ′ (white). Upon observing Fig. 6, it can be noted that there are no significant differences in the size of the γ′ phase among the aged CNS1, CNS2, and CNS3 samples. However, the γ′ area fraction varied between approximately 43.58±7.8% for CNS3 to 68.45±0.98% and 68.82±1.5% for CNS1 and CNS2, respectively. The distinct morphology of γ′ observed in the aged samples (Fig. 6) can be attributed to variations in element partitioning between the γ and γ′ phases. Furthermore, the deviation in γ′ shape observed in CNS2 and CNS3 samples (Figs. 6b and c) compared to the cubic morphology in CNS1 (Fig. 6a) indicates a change in the misfit value between γ and γ′ phases. The underlying reasons for these distinct microstructural features in samples with different chemical compositions will be discussed in the following sections.

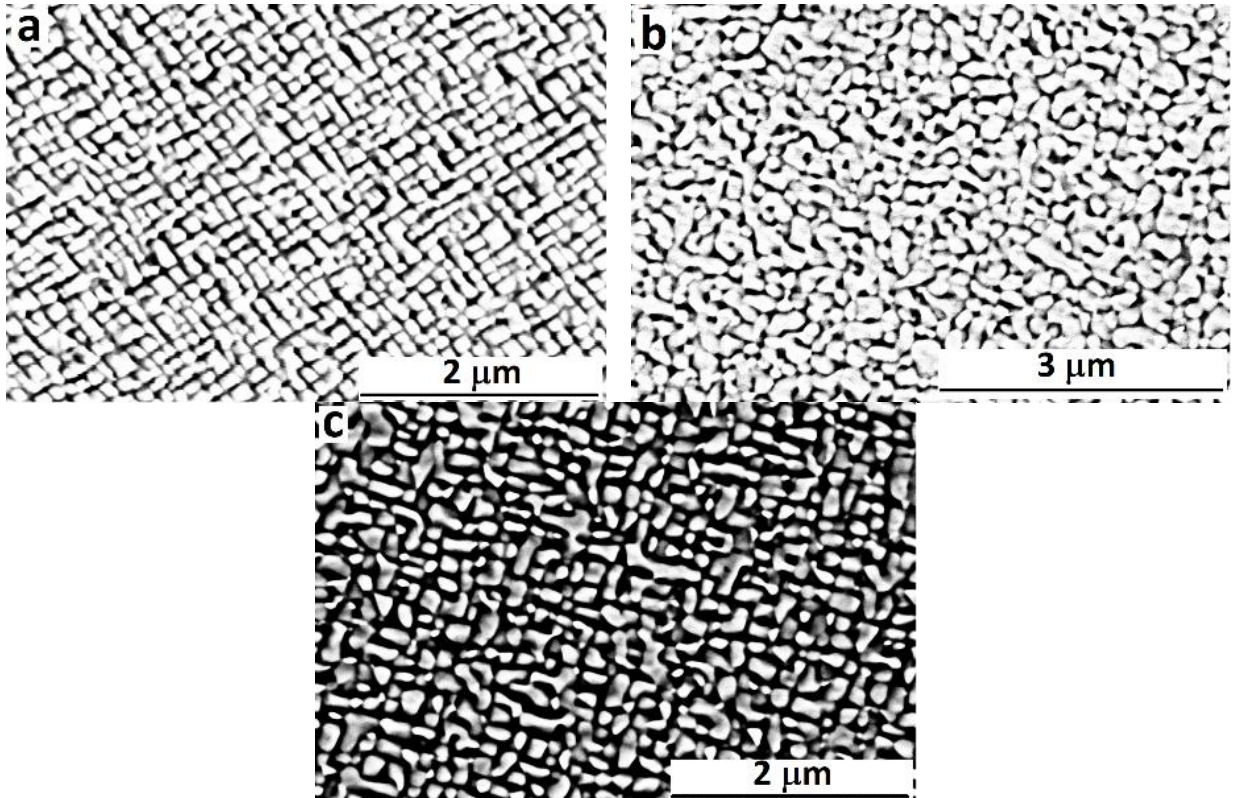

Fig. 6. SEM micrographs of the samples aged at 900 °C for 24 hours, followed by water quenching: (a) CNS1, (b) CNS2, and (c) CNS3. The black and white areas represent the presence of the γ and γ' phases, respectively.

Fig. 7 illustrates a representative grain boundary micrograph and related chemical analyses obtained from the CNS1 sample. The sample underwent homogenization at 1250 °C, followed by air cooling, and aging at 900 °C, followed by water quenching. The SEM back-scattered image reveals a black particle and two types of precipitations at the grain boundary. The black particle, indicated by a white arrow, corresponds to an Al oxide core surrounded by a Ti nitride shell, as confirmed by mapping analysis (yellow arrows). Based on the CALPHAD results for the CNS1 alloy, the presence of μ phase at the grain boundary is more possible. The EDS mapping results indicate that the discontinuous white particles, encircled in red, are likely the μ phase enriched with Ti, W, and Ta. Additionally, gray areas at the grain boundaries, highlighted by dashed rectangles, exhibit enrichment of Co and Cr and represent the γ' precipitate-depleted zones (PDZ), predominantly appearing near the TCP phases [27,40,52]. It is worth mentioning that in the case of CNS2 and CNS3 alloys, a small amount of μ phase with PDZ at the grain boundaries was also observed (see supplementary material, **S2** and **S3**).

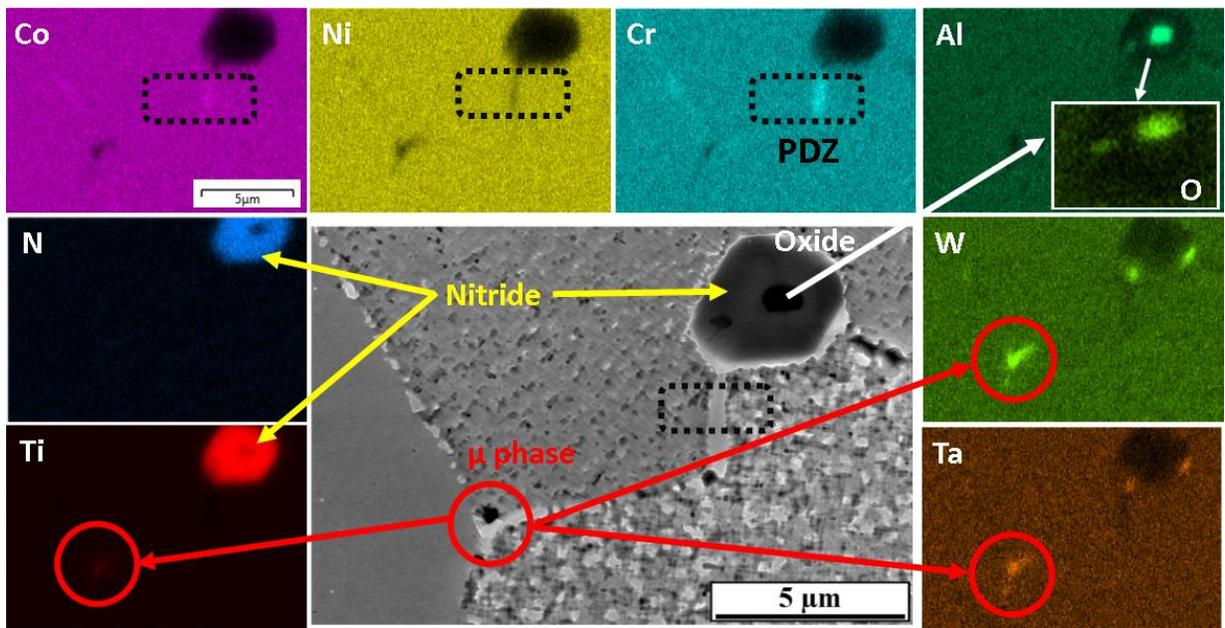

Fig. 7. A representative back-scattered SEM micrograph of grain boundaries and corresponding EDS maps in the CNS1 alloy after homogenization at 1250 °C, followed by air cooling, and aging at 900 °C, followed by water quenching.

*3.4 Mechanical properties*

Nano-indentation maps were utilized to verify the effectiveness of the homogenization treatment on the as-cast alloys. Fig. 8 presents the nano-hardness and reduced modulus maps of the as-cast and homogenized CNS3 alloy. In the vicinity of the β phase in the as-cast alloy (Figs. 8a and b), higher nano-hardness values and lower reduced modulus values were observed compared to other areas of the alloy. However, after the homogenization process, a uniform distribution of nano-hardness and reduced modulus was achieved (Fig. 8c and d), confirming the dissolution of the β phase.

The micro-hardness values of the developed alloys in their various as-cast, homogenized, and aged conditions are plotted in Fig. 9a for comparison. It is evident that the CNS1 alloy exhibited higher hardness values than the other alloys in all conditions. Furthermore, the homogenization process led to a reduction in the hardness of the alloys. Subsequently, after the aging process, the alloys reached their maximum hardness.

Fig. 9b presents the load-depth curve, hardness, and reduced modulus values obtained from the nano-indentation data. As shown in the figure, the CNS1 alloy exhibited the highest hardness (consistent with the micro-hardness results) and reduced modulus among the samples, while the CNS2 alloy displayed the lowest corresponding values. These hardness values align well with the available literature [27,40]. It should be noted that the difference in hardness values between the γ and γ' phases could not be directly measured due to the small size of the γ' phase. Therefore, all hardness values represent the average hardness across the microstructure for each respective alloy.

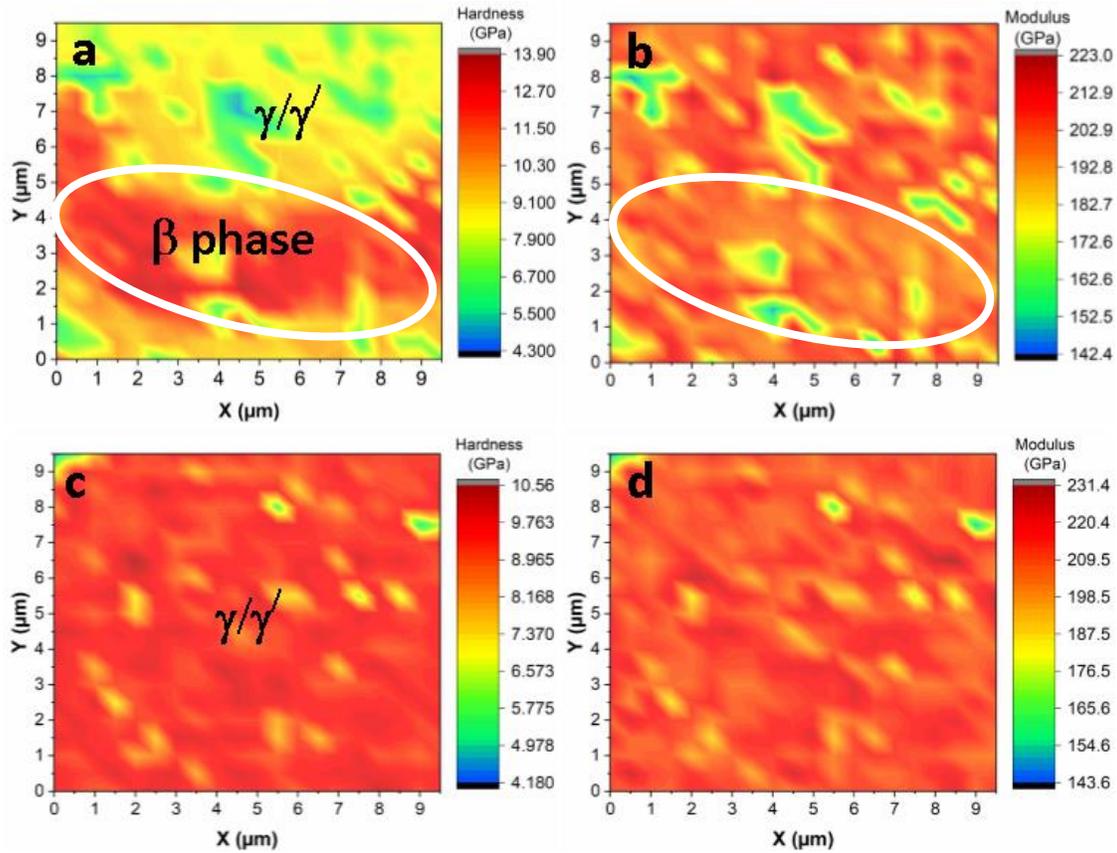

Fig. 8. Nano-hardness and reduced modulus maps of the CNS3 sample obtained from nano-indentation test: (a) as-cast-alloy nano-hardness map, (b) as-cast-alloy reduced modulus map, (c) homogenized condition nano-hardness map, and (d) homogenized condition reduced modulus map. The white ellipse in (a) and (b) refers to the vicinity of the β phase with higher hardness and lower modulus.

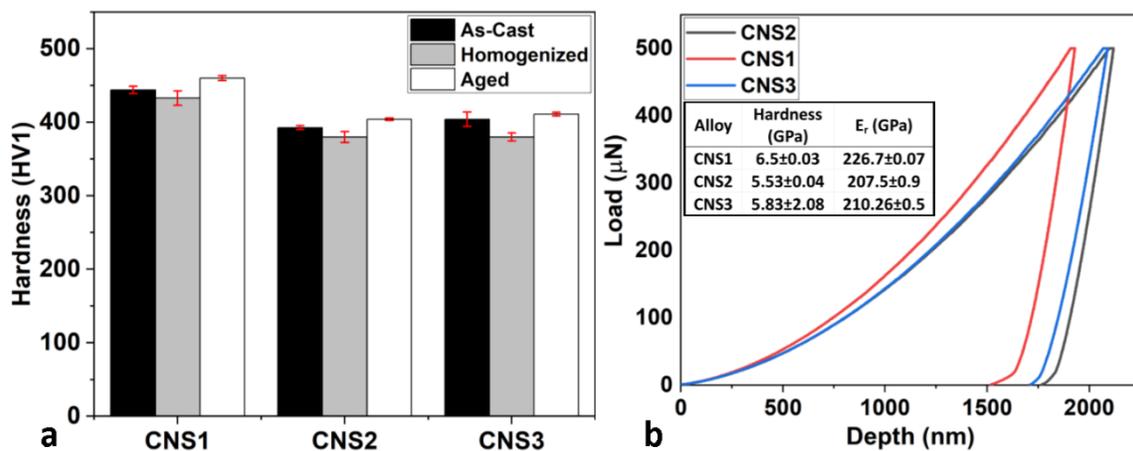

Fig. 9. (a) Micro-hardness distribution of the developed alloys in their as-cast, homogenized, and aged conditions, (b) the load-depth plot of the aged samples shows hardness and reduced modulus values obtained from the nano-indentation test.

# 4 Discussion

## 4.1 Effect of $\Delta S^{mix}$

High-angle annular dark field (HAADF) STEM image, bright-field TEM images, and corresponding STEM-EDS maps of different elements in alloys CNS1, CNS2, and CNS3, which were homogenized at 1250 °C for 24 hours, followed by air cooling and aged at 900 °C for 24 hours, followed by water quenching, are shown in Figs. 10 and 11. Fig. 12 presents the plot of the γ′/γ partitioning coefficient ($K_{\gamma'/\gamma}$) for various elements, calculated from STEM-EDS line scan analysis. Table 4 summarizes the chemical composition and $K_{\gamma'/\gamma}$ values.

According to $K_{\gamma'/\gamma}^{Co}$ and $K_{\gamma'/\gamma}^{Cr}$, Co and Cr (γ stabilizers) preferentially partition into the γ phase, which is consistent with the literature [45,53]. Figs. 10-11 indicate that Ti, Ta, Al, Ni, and W (γ′ stabilizers) tend to be partitioned into the γ′ phase. However, the partitioning of W into the γ′ phase (with $K_{\gamma'/\gamma}^{W}$ values of 1.05, 1.02, and 1.04 for CNS1, CNS2, and CNS3 alloys, respectively) is not as pronounced as that of Ti, Ta, Al, and Ni. The partitioning of Ni into the γ′ phase widens the γ+ γ′ region, offering a broader heat treatment window [17,54].

In terms of γ and γ′ formers, it can be observed that W and V have equal contributions. While W and Cr are known as solid solution-strengthening elements [23], in this case, they resulted in a rounded morphology for γ′ (CNS2 and CNS3) instead of the desired cuboidal shape. It is important to note that the addition of W and Cr can lead to the precipitation of detrimental phases. However, in a system with a higher $\Delta S_{mix}$ value, sluggish diffusion [55] can impede the precipitation of these intermetallics. This suggests that increasing the Cr content further might improve oxidation resistance and promote a cuboidal γ′ morphology while avoiding the formation of detrimental intermetallics.

However, based on the $K_{\gamma'/\gamma}^{V}$ values of 0.95 and 0.92 for CNS2 and CNS3, respectively, and the STEM-EDS maps for V shown in Fig. 11, it can be observed that V has a neutral effect on the amount of γ and γ′ phases, which has not been reported in the literature.

A comparison of the chemical compositions of CNS1 and CNS2 alloys (Table 1) reveals a reduction in the amount of Cr from 13 to 9 at. % and an increase in the amount of V from 0 to 4 at. %. The variations in $K_{\gamma'/\gamma}$ for CNS1 and CNS2 alloys (Fig. 12) indicate that the addition of V and reduction of Cr led to a decrease in $K_{\gamma'/\gamma}^{Ni}$ (from 1.61 to 1.48), and an increase in both $K_{\gamma'/\gamma}^{Al}$ (from 1.57 to 1.77) and $K_{\gamma'/\gamma}^{Co}$ (from 0.77 to 0.82) in CNS2 compared to CNS1. Despite the γ stabilizing effect of Cr and the neutral effect of V (Figs. 10-12), the amount of γ′ phase remains unchanged (approximately 68% for CNS1 and CNS2). This indicates the presence of a positive metallurgical parameter compensating for the negative effect of Cr on the γ′ area fraction. In fact, increasing the nominal ΔS$_{mix}$ value from 1.488R to 1.568R through the addition of V at the expense of Cr stabilizes the γ′ phase and enhances the T$_{\gamma'}$ from 1150 to 1165 °C (Fig. 5 and Table 3).

The developed alloys in this study are based on well-known CoNi-based superalloys recommended by Lass [29] and Zhuang et al. [27]. The ΔS$_{mix}$ and T$_{\gamma'}$ values for the developed alloy in [27] are 1.457R and 1099 °C, respectively. By increasing the entropy from 1.457R to 1.494R in the CNS1 alloy, T$_{\gamma'}$ increased from 1099 °C to 1150 °C (approximately 55 °C). In terms of *sustainability*, especially in the case of superalloys, it is crucial to apply more selection and design procedures (pre-build strategies) to enhance the high-temperature service capability, which directly depends on the γ′ solvus temperature.

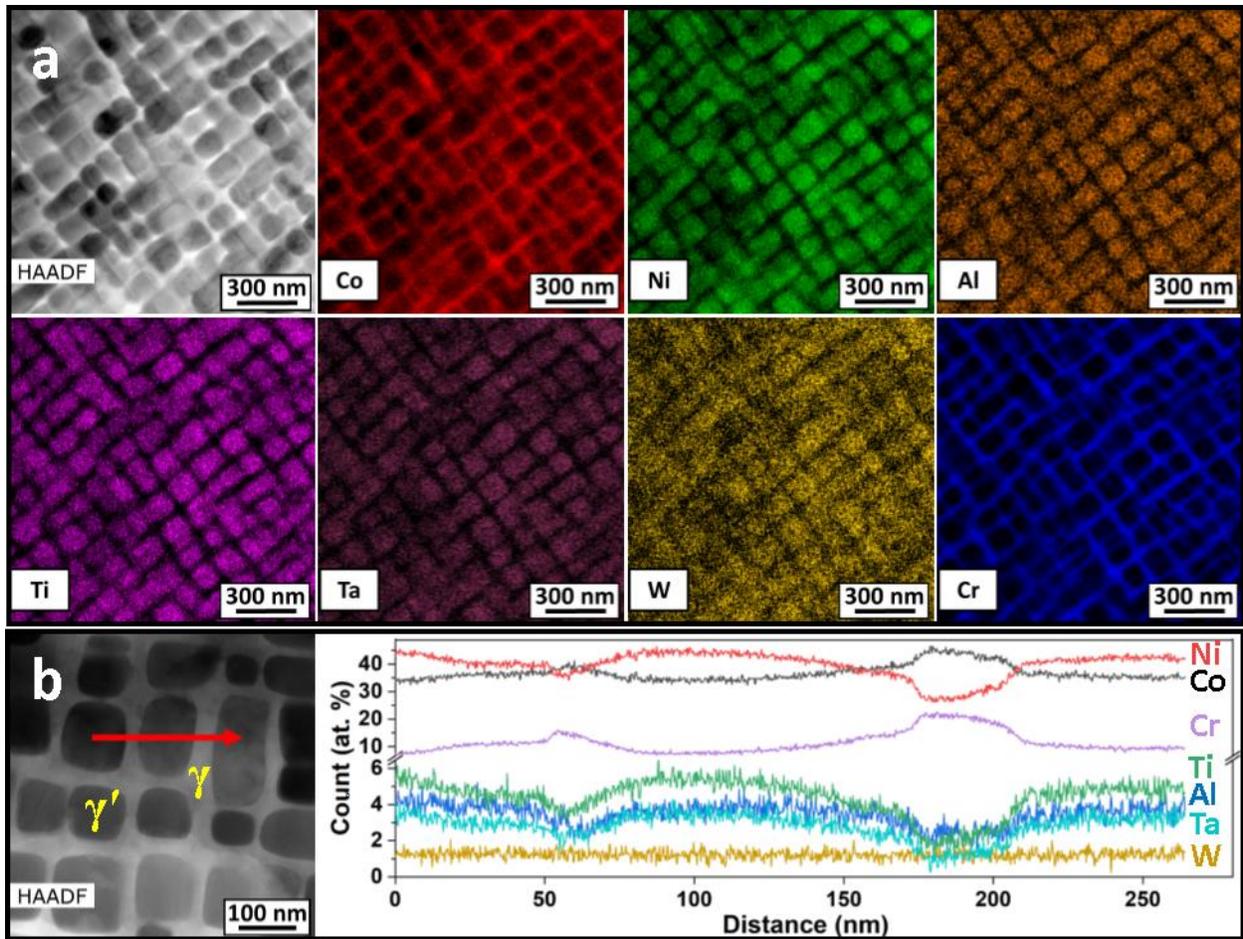

Fig. 10. (a) HAADF STEM image and corresponding STEM-EDS maps of different elements and (b) a typical STEM-EDS line scan from γ′ and γ phases of alloy CNS1 aged at 900 °C for 24 h followed by water quenching.

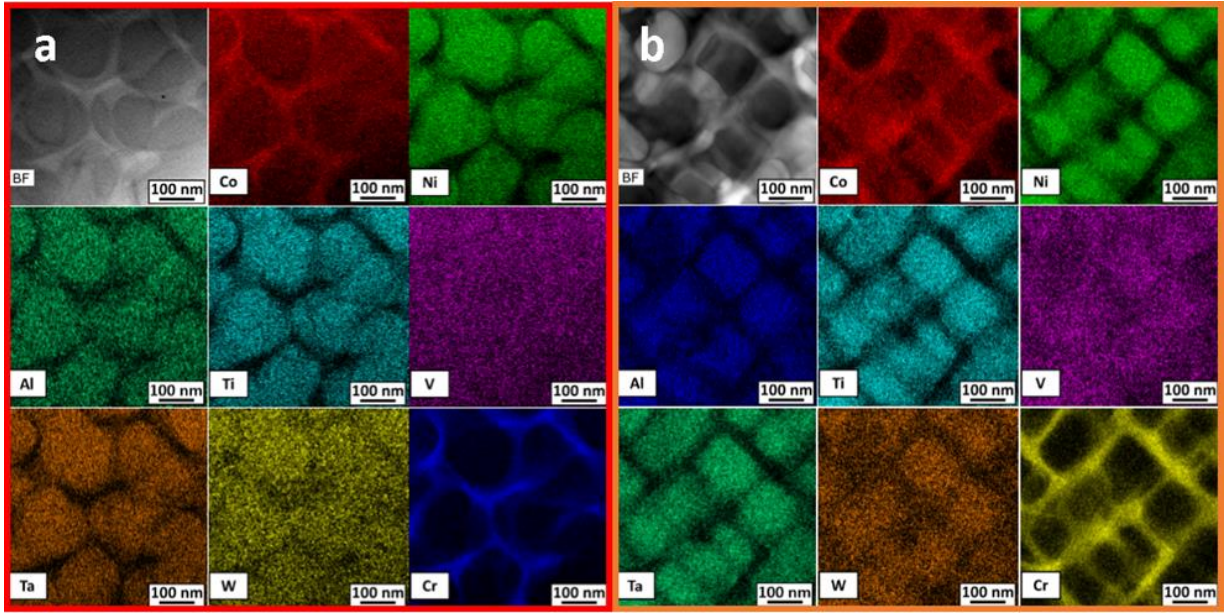

Fig. 11. (a) BF TEM images and corresponding STEM-EDS maps of different elements of alloys CNS2 and CNS3 homogenized at 1250 °C for 24h followed by air cooling and aged at 900 °C for 24 h followed by water quenching.

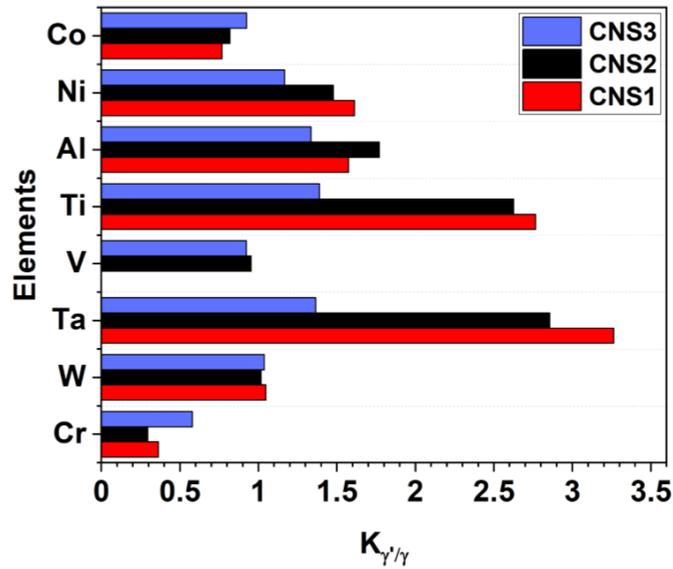

Fig. 12. γ′/γ partitioning coefficients obtained from STEM-EDS analysis of alloys CNS1, CNS2, and CNS3 after homogenization at 1250 °C for 24h followed by air cooling and aging at 900 °C for 24h followed by water quenching.

Table 4. Chemical composition and $K_{\gamma'/\gamma}$ of the γ and γ′ phases in different alloys. $K_{\gamma'/\gamma}$ was obtained from STEM-EDS line scans by averaging corresponding thirty points.

|  |  | Chemical composition (at. %) | | | | | | | |
|---|---|---|---|---|---|---|---|---|---|
|  |  | Co | Ni | Al | Ti | V | Ta | W | Cr |
| γ′ | CNS1 | 34.26 | 44.13 | 3.73 | 5.40 | 0.00 | 3.48 | 1.26 | 7.73 |
|  | CNS2 | 35.71 | 41.79 | 4.52 | 4.15 | 3.70 | 3.52 | 1.49 | 5.13 |
|  | CNS3 | 39.15 | 36.03 | 5.74 | 4.97 | 1.86 | 3.05 | 1.55 | 7.66 |
| γ | CNS1 | 44.67 | 27.37 | 2.37 | 1.95 | 0.00 | 1.07 | 1.20 | 21.36 |
|  | CNS2 | 43.58 | 28.31 | 2.55 | 1.58 | 3.88 | 1.23 | 1.46 | 17.39 |
|  | CNS3 | 42.29 | 30.87 | 4.30 | 3.58 | 2.01 | 2.23 | 1.50 | 13.22 |
| $K_{\gamma'/\gamma}$ | CNS1 | 0.77 | 1.61 | 1.57 | 2.77 | 0.00 | 3.26 | 1.05 | 0.36 |
|  | CNS2 | 0.82 | 1.48 | 1.77 | 2.62 | 0.95 | 2.86 | 1.02 | 0.29 |
|  | CNS3 | 0.93 | 1.17 | 1.33 | 1.39 | 0.92 | 1.36 | 1.04 | 0.58 |

*4.2 Importance of γ′ stabilizers*

To investigate the impact of γ′ stabilizers, the Ni content in CNS3 was reduced from 35 to 30 at. %, while the V content decreased from 4 to 2 at. % and the Co and Cr contents increased from 36 to 40 at. % and 9 to 12 at. %, respectively. This composition was chosen to maintain a relatively constant $\Delta S_{mix}$ in CNS2 and CNS3 (Table 2). A comparison of the microstructural features of CNS2 and CNS3 alloys (Figs. 6 and 11b) reveals that despite the higher $\Delta S_{mix}$ value, the amount of γ′ phase decreased from approximately 68% in CNS2 to about 44% in CNS3. Additionally, the $T_{\gamma'}$ decreased from 1165 to 1134 °C (Fig. 5 and Table 3). These observations highlight the significance of $\Delta S_{mix}$ and the quantity of γ′ stabilizers in determining the fraction of the γ′ phase and its corresponding $T_{\gamma'}$. Therefore, when designing alloys, careful consideration of the amount of γ′ stabilizing elements, in conjunction with the high entropy concept, is advised. Furthermore, it is important to note that there are optimal amounts of Ti and Ta to prevent the coarsening of γ′ precipitates and the formation of TCP phases during aging. It is recommended to limit the Ti content to a maximum of 4 at. % [31], while the Ta content should be kept below 3 at. % to avoid the occurrence of TCP phases [29]. Fig. 13 presents a comparison of the $T_{\gamma'}$ values of the developed alloys in this study and other high Cr CoNi-based superalloys with respect to $\Delta S_{mix}$. It is evident that the developed alloys exhibit significantly higher $T_{\gamma'}$ compared to those reported in the literature [27,34,35,52,54,56]. This further confirms that both the type and quantity of elements have an impact on the final $T_{\gamma'}$. A similar approach has been employed in the development of Ni-based superalloys, such as the sixth-generation superalloy TMS238, which possesses superior high-temperature properties due to a $\Delta S_{mix}$ of 1.68R, resulting from a multi-element alloying system that includes refractory elements such as Re and Ru [25].

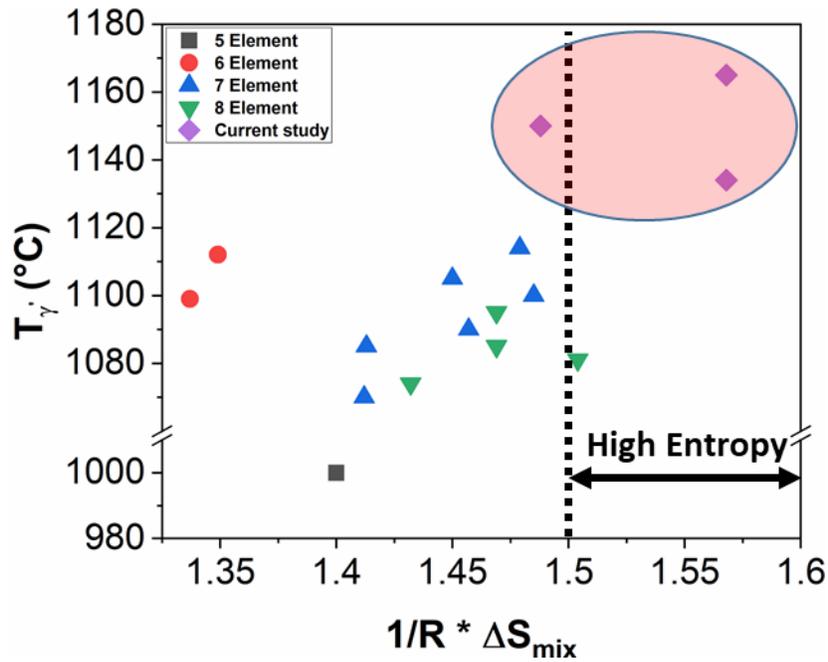

**Fig. 13.** Effect of the number of elements and $\Delta S_{mix}$ on $T_{\gamma'}$ in high Cr CoNi superalloys. Data are collected from the literature [27,34,35,52,54,56].

*4.3 γ' morphology*

Elastic stress and interfacial energy are two influential experimental parameters that determine the final morphology of precipitates, specifically the γ' phase in superalloys. The morphology tends to be cubic when elastic stress, dependent on the lattice misfit between γ and γ' phases, is the predominant factor. Conversely, when the lattice misfit or elastic stress is low, the interfacial energy becomes the governing parameter during the coarsening of the γ' phase [57–59].

In this study, the addition of Cr to the developed alloys aims to enhance oxidation resistance. However, Cr also tends to partition into the γ phase (Figs. 10-12), which can reduce the lattice misfit. Interestingly, despite the high Cr content (13 at. %) in the CNS1 alloy, the γ' phase maintains a cubic morphology (Fig. 10). This suggests that the alloy design based on the high configurational entropy concept ($\Delta S_{mix}$ = 1.494R in CNS1 alloy) compensates for the negative effect of Cr on lattice misfit, allowing the γ' phase to retain its cubic morphology. It is worth noting that the morphology and volume fraction of the γ' phase significantly impact the final hardness of the superalloys [45]. From Fig. 9, it can be observed that the CNS1 alloy exhibits the highest hardness values compared to CNS2 and CNS3, thanks to its cubic morphology and high volume fraction of the γ' phase. However, CNS1 has a lower $T_{γ'}$ compared to CNS2 (Fig. 5 and Table 3).

In the case of CNS2, despite the lower Cr content (reduced lattice misfit) and higher $\Delta S_{mix}$ (increased lattice misfit) compared to CNS1, the γ' regions have transitioned from a cubic to a rounded morphology (Figs. 10 and 11). V tends to have a uniform distribution within the γ and γ' phases (Figs. 10-12), indicating its limited effect on lattice misfit. Therefore, the possible mechanism is the positive effect of V on interfacial energy, leading to the formation of a more rounded γ' phase. Furthermore, Figs. 10 and 11 show higher concentrations of γ' stabilizing elements (such as Ti, Ta, Al, and Ni) in the interfacial area between γ and γ' phases, suggesting that V hinders their diffusion into the γ' phase. This hindrance can negatively affect the lattice misfit. Overall, since $K^V_{γ'/γ}$ is approximately equal to 1, V indirectly influences interfacial energy and lattice misfit by affecting the diffusion of other elements. In CNS3, compared to CNS2, higher amounts of Co and Cr (with a high partitioning tendency to the γ phase, reducing lattice misfit) and a lower amount of Ni (a strong γ' stabilizer with high partitioning tendency to the γ' phase, reducing lattice misfit) lead to the dominance of interfacial energy. As a result, the morphology of the γ' phase tends to be more rounded than cubic (Fig. 6).

## 5 Conclusions

To enhance the high-temperature performance of high Cr CoNi-based superalloys, different alloys were designed using CALPHAD calculations based on the concept of high configurational entropy for future sustainable manufacturing technologies. The following conclusions can be summarized:

i. Thermodynamic CALPHAD calculations predict the final microstructures of the designed CoNi-based HESAs and transformation temperatures within the explored range of compositions with good accuracy. Hence, in terms of predicting $T_S$ and $T_{\gamma'}$ the results are not well aligning with experimental data. After casting, the microstructures consist of an FCC matrix with microsegregation and some β phase. Homogenization treatment at 1250 °C for 24 hours followed by air-cooling results in a homogenized and more suitable microstructure for aging compared to furnace-cooling. After aging, the developed alloys exhibit different area fractions and morphologies of γ′ within the γ matrix.

ii. Increasing the $\Delta S_{mix}$ can compensate for the negative effect of γ stabilizers such as Cr, and this can be achieved by adding a neutral alloying element like V instead. Therefore, substituting V for Cr to increase $\Delta S_{mix}$ stabilizes the γ′ phase and also enhances the γ′ solvus temperature.

iii. Besides $\Delta S_{mix}$, the choice of alloying elements plays a crucial role in determining the final γ′ solvus temperature and volume fraction. During alloy design, the amount of γ′ stabilizers should not be reduced in order to increase $\Delta S_{mix}$. Therefore, a promising strategy to enhance high-temperature characteristics is to increase $\Delta S_{mix}$ while simultaneously increasing the amount of γ′ stabilizers in the chemical composition.

iv. By employing the presented strategy, high Cr CoNi-based HESAs can be produced with a high γ' solvus temperature (1165 °C), a processing window of 120 °C to 141 °C, a freezing range of 61 °C to 71 °C, and hardness values comparable to recently developed CoNi-based superalloys. These characteristics make the developed alloys promising materials for sustainable manufacturing technologies, such as powder bed fusion technologies, which represent a significant topic for future research.

**CRediT authorship contribution statement**

**Ahad Mohammadzadeh**: Conceptualization, Methodology, Data and experiments curation and analysis, Funding and Investigation, Writing-original draft, review & editing. **Akbar Heidarzadeh**: Conceptualization, Methodology, Investigation, Writing-review & editing. **Hailey Becker**: TEM and nanoindentation study as a part of her internship project, Writing-review & editing. **Jorge Valilla Robles**: CALPHAD study, review & editing. **Alberto Meza**: Investigation, sample preparation, casting, review & editing. **Manuel Avella**: TEM test, review & editing. **Miguel Monclús**: nanoindentation test, review & editing. **Damien Tourret**: CALPHAD study, Writing-review & editing. **Jose M. Torralba**: Conceptualization, Resources, Project administration, Writing-review & editing, Supervision.

**Declaration of competing interest**

The authors declare no conflict of interest.

**Data availability statement**

The raw/processed data required to reproduce these findings cannot be shared at this time as the data also forms part of an ongoing study.

**Acknowledgment**

This investigation was supported by the European Union Horizon 2020 research and innovation programme (Marie Sklodowska-Curie Individual Fellowships, Grant Agreement 101028155). DT gratefully acknowledges support from the Spanish Ministry of Science through a Ramón y Cajal Fellowship (Ref. RYC2019-028233-I). Also, the National Science Foundation Division of Material Research (Grant No. DMR2153316) through the International Research Experience for Students (IRES) program supported a portion of the funding for this research. Amalia Sanromán and Ignacio Escobar are sincerely thanked for their help with VAR processing and sample preparations.